\documentclass[a4paper,12pt]{article}
\usepackage{cite}
\newcommand{\be}{\begin{equation}}
\newcommand{\ee}{\end{equation}}
\newcommand{\bea}{\begin{eqnarray}}
\newcommand{\nn}{\nonumber}
\newcommand{\eea}{\end{eqnarray}}

\begin{document}

\begin{titlepage}

\begin{flushright}
UB-ECM-PF-04/14\\
\end{flushright}
\begin{centering}
\vspace{.3in}
{\large{\bf Self-Gravitational Corrections to the Cardy-Verlinde Formula and the FRW Brane Cosmology in $SdS_5$ Bulk
}}

\vspace{.5in} {\bf Mohammad R. Setare$^{1}$ and Elias C.
Vagenas$^{2}$ }\\
\vspace{.3in} $^{1}$\,Physics Dept., Inst. for Studies in Theo. Physics and
Mathematics (IPM)\\
P.O. Box 19395-5531, Tehran, Iran\\
rezakord@ipm.ir\\
\vspace{0.4in}

$^{2}$\, Departament d'Estructura i Constituents de la Mat\`{e}ria\\
and\\ CER for Astrophysics, Particle Physics and Cosmology\\
Universitat de Barcelona\\
Av. Diagonal 647, E-08028 Barcelona\\
Spain\\
evagenas@ecm.ub.es\\
\end{centering}

\vspace{0.7in}
%%%%%%%%%%%%%%%%%%%ABSTRACT%%%%%%%%%%%%%%%%%%%%%%%%%%%%%%%%%%%%%%%%%%%%%%%%%%%%%%%%%%%%%%%%%%%%%%
%%%%%%%%%%%%%%%%%%%%%%%%%%%%%%%%%%%%%%%%%%%%%%%%%%%%%%%%%%%%%%%%%%%%%%%%%%%%%%%%%%%%%%%%%%%%%%%%%
%%%%%%%%%%%%%%%%%%%%%%%%%%%%%%%%%%%%%%%%%%%%%%%%%%%%%%%%%%%%%%%%%%%%%%%%%%%%%%%%%%%%%%%%%%%%%%%%%
%%%%%%%%%%%%%%%%%%%%%%%%%%%%%%%%%%%%%%%%%%%%%%%%%%%%%%%%%%%%%%%%%%%%%%%%%%%%%%%%%%%%%%%%%%%%%%%%%
%%%%%%%%%%%%%%%%%%%%%%%%%%%%%%%%%%%%%%%%%%%%%%%%%%%%%%%%%%%%%%%%%%%%%%%%%%%%%%%%%%%%%%%%%%%%%%%%%

\begin{abstract}
%%%%%%%%%%%%%%%%

The semiclassical corrections to the Cardy-Verlinde entropy  of a five-dimensional Schwarzschild de-Sitter black
hole ($SdS_{5}$) are explicitly evaluated. These corrections are considered within the context of KKW analysis and
arise as a result of the self-gravitation effect. In addition, a four-dimensional spacelike brane is considered as
the boundary of the $SdS_{5}$ bulk background. It is already known that the induced geometry of the brane is
exactly given by that of a radiation-dominated FRW universe. By exploiting the CFT/FRW-cosmology relation, we derive
the self-gravitational corrections to the first Friedmann-like equation which is the equation of the brane motion.
The additional term that arises due to the semiclassical analysis can be viewed as stiff matter where the
self-gravitational corrections act as the source for it. This result is contrary to standard analysis that regards
the charge of $SdS_{5}$ bulk black hole as the source for stiff matter. Furthermore, we rewrite the Friedmann-like
equation in a such way that it represents the conservation equation of energy of a point particle
moving in a one-dimensional effective potential. The self-gravitational corrections
to the effective potential and, consequently, to the point particle's motion are obtained.
A short analysis on the asymptotic behavior of the $4$-dimensional brane is presented.

%%%%%%%%%%%%%%%%
\end{abstract}

%%%%%%%%%%%%%%%%%%%%%%%%%%%%%%%%%%%%%%%%
%%%%%%%%%%%%%%%%%%%%%%%%%%%%%%%%%%%%%%%%
%%%%%%%%%%%%%%%%%%%%%%%%%%%%%%%%%%%%%%%%
\end{titlepage}
\newpage
\baselineskip=18pt
%%%%%%%%%%%%%%%%%%%%%%%%%%%%%%%%%%%%%%%%%%%%%%%%%%%%%%%%%%%%%%%%%%%%%%%%%%%%%%%%%%%%%%%%%%%%%%%%%
%%%%%%%%%%%%%%%%%%%%%%%%%%%%%%%%%%%%%%%%%%%%%%%%%%%%%%%%%%%%%%%%%%%%%%%%%%%%%%%%%%%%%%%%%%%%%%%%%
%%%%%%%%%%%%%%%%%%%%%%%%%%%%%%%%%%%%%%%%%%%%%%%%%%%%%%%%%%%%%%%%%%%%%%%%%%%%%%%%%%%%%%%%%%%%%%%%%
%%%%%%%%%%%%%%%%%%%%%%% Introduction%%%%%%%%%%%%%%%%%%%%%%%%%%%%%%%%%%%%%%%%%%%%%%%%%%%%%%%%%%%%%
\section{Introduction}
Concerning the quantum process called Hawking effect \cite{hawking1} much work has been done using a fixed
background during the emission process. The idea of Keski-Vakkuri, Kraus and Wilczek (KKW) \cite{KKW} is to view
the black hole background as dynamical by treating the Hawking radiation as a tunnelling process. The energy
conservation is the key to this description. The total (ADM) mass is kept fixed while the mass of the black hole
under consideration decreases due to the emitted radiation. The effect of this modification gives rise to
additional terms in the formulae concerning the known results for black holes \cite{corrections}; a non-thermal
partner to the thermal spectrum of the Hawking radiation shows up.

\par\noindent
Recent astronomical observations of supernovae and cosmic microwave background \cite{Ries98} indicate that the
universe is accelerating and can be well approximated by a world with a positive cosmological constant. If the
universe accelerates indefinitely, the standard cosmology leads to an asymptotic dS universe. In addition, de
Sitter spacetime plays an important role in the inflationary scenario where an exponentially expanding
approximately dS spacetime is employed to solve a number of problems in standard cosmology. Furthermore, the
quantum field theory on dS spacetime is also of considerable interest.

\par\noindent
Much attention has currently been paid for the duality between de Sitter (dS) gravity and CFT in  analogy to the
AdS/CFT correspondence \cite{ds/cft}(for a very good review see also \cite{odi}).
The reason is that the isometry of ($n+1$)-dimensional de Sitter space, $SO(n+1,1)$,
exactly agrees with the conformal symmetry of $n$-dimensional Euclidean space.
Thus, it might be natural to expect the correspondence between ($n+1$)-dimensional gravity in de
Sitter space and $n$-dimensional Euclidean CFT (the dS/CFT correspondence). Moreover, the holographic principle
between the radiation dominated Friedmann-Robertson-Walker (FRW) universe in $n$-dimensions and same dimensional
CFT with a dual ($n+1$)-dimensional AdS description was studied in \cite{verlinde}. Particularly, one can see the
correspondence between black hole entropy and the entropy of the CFT which is derived by making the appropriate
identifications for the Friedmann equation with the  Cardy-Verlinde formula.

\par\noindent
There has been much recent interest in calculating the quantum corrections to $S_{BH}$ (the Bekenstein-Hawking
entropy) \cite{logcorrections1}. The leading-order correction is proportional to $\ln{S_{BH}}$. In a recent work,
Carlip \cite{carlip1} has deduced the leading order quantum correction to the classical Cardy formula. The Cardy
formula follows from a saddle-point approximation of the partition function for a two-dimensional conformal field
theory. This leads to the theory's density of states, which is related to the partition function by way of a
Fourier transform \cite{carlip2}.

\par\noindent
In the present paper, we take into account corrections to the entropy of the  five-dimensional Schwarzschild-de
Sitter black hole (abbreviated to $SdS_5$ in the sequel) that arise due to the self-gravitational effect. Previous
studies of the Cardy-Verlinde formula (or the corresponding Friedmann equation) in an dS/CFT context have thus far
neglected this sub-dominant, but important, contribution\footnote{On the contrary, self-gravitational corrections
to thermodynamical quantities of  AdS spaces have attracted a lot of attention. For instance, the authors have
shown that the entropy of Ach\'ucarro-Ortiz black hole (a locally AdS space) can be described by the
Cardy-Verlinde formula \cite{setelias1} and the self-gravitational corrections to this formula have been computed
\cite{setelias2}.}.
%%%%%%%%%%%%%%%%%%%%%%%%%%%%%%%%%%%%%%%%%%%%%%%%%%%%%%%%%%%%%%%%%%%%%%%%%%%%%%%%%%%%%%%%%%%%%%%%%
%%%%%%%%%%%%%%%%%%%%%%%%%%%%%%%%%%%%%%%%%%%%%%%%%%%%%%%%%%%%%%%%%%%%%%%%%%%%%%%%%%%%%%%%%%%%%%%%%
%%%%%%%%%%%%%%%%%%%%%%%%%%%%%%%%%%%%%%%%%%%%%%%%%%%%%%%%%%%%%%%%%%%%%%%%%%%%%%%%%%%%%%%%%%%%%%%%%
%%%%%%%%%%%%%%%%%%%%%%%%%%%%%%%%%%%%%%%%%%%%%%%%%%%%%%%%%%%%%%%%%%%%%%%%%%%%%%%%%%%%%%%%%%%%%%%%%
\section{Schwarzschild-de Sitter black hole}
The $SdS_5$ black hole is a constant curvature solution of the Einstein equation which follows from the action
\bea
S=\int d^{5}x \sqrt{-g}\left[ \frac{R}{16\pi G_5}+\Lambda_{5}\right]
\hspace{1ex},
\label{act}
\eea
here $R$ is the scalar
curvature, $\Lambda_{5}$ is the five-dimensional positive cosmological constant, and $G_5$ denotes the
five-dimensional Newton's constant. The corresponding line element is given by
\bea
ds^{2}=-\left(1-\frac{r^{2}}{l^{2}}-\frac{M\varepsilon_{3}}{r^{2}}\right)
dt^{2}+\left(1-\frac{r^{2}}{l^{2}}-\frac{M\varepsilon_{3}}{r^{2}}\right)^{-1}dr^{2}+r^{2} d\Omega_{(3)}^{2}
\hspace{1ex},
\label{metric1}
\eea
where $\varepsilon_3=\frac{16\pi G_5}{3V_3}$, $M$ is the black hole mass \cite {BBM},  $V_3$ is the volume of
the spherical hypersurface described by $d\Omega_{3}^{2}$, and $l$ represents the curvature radius of the $SdS_5$
bulk space. The cosmological constant and the curvature radius are related by $\Lambda_{5}=\frac{6}{l^{2}}$. The
$SdS_5$ cosmological horizon is described by the largest root of
\bea
\frac{r^{4}}{l^{2}}- r^{2}+M\varepsilon_3=0
\label{larg}
\eea
and thus it takes the form
\bea
r_{c}^{2}=\frac{l^{2}}{2}\left(1+\sqrt{1-\frac{4M\varepsilon_3}{l^{2}}}\right)
\hspace{1ex},
\label{coshor}
\eea
 while the black
hole horizon takes the form \bea r_{bh}^{2}=\frac{l^{2}}{2}\left(1-\sqrt{1-\frac{4M\varepsilon_3}{l^{2}}}\right)
\label{eventhor} \hspace{1ex}. \eea It is evident that due to the presence of two horizons in $SdS_5$ black hole
spacetime, one can define two different Hawking temperatures. The temperature of the cosmological horizon
($r_{c}$) is given as \bea T_{c}=-\left(\frac{1}{2\pi r_{c}}-\frac{r_{c}}{\pi l^{2}}\right) \label{costem} \eea
while the temperature of black hole horizon ($r_{bh}$) reads \bea T_{bh}=\frac{1}{2\pi r_{bh}}-\frac{r_{bh}}{\pi
l^{2}} \hspace{1ex}. \label{blatem} \eea The entropy $S$ for both horizons is \bea S=\frac{V_3 r_{H}^{3}}{4G_5}
\label{entro} \eea where $r_{H}$ takes the values $r_{c}$ and $r_{bh}$ for the cosmological and the black hole
horizons, respectively. The thermodynamical energy is given by \bea E=\pm M \label{ener} \eea where $+$
corresponds to the black hole horizon and $-$ to the cosmological horizon.
%%%%%%%%%%%%%%%%%%%%%%%%%%%%%%%%%%%%%%%%%%%%%%%%%%%%%%%%%%%%%%%%%%%%%%%%%%%%%%%%%%%%%%%%%%%%%%%%
%%%%%%%%%%%%%%%%%%%%%%%%%%%%%%%%%%%%%%%%%%%%%%%%%%%%%%%%%%%%%%%%%%%%%%%%%%%%%%%%%%%%%%%%%%%%%%%%
%%%%%%%%%%%%%%%%%%%%%%%%%%%%%%%%%%%%%%%%%%%%%%%%%%%%%%%%%%%%%%%%%%%%%%%%%%%%%%%%%%%%%%%%%%%%%%%%
%%%%%%%%%%%%%%%%%%%%%%%%%%%%%%%%%%%%%%%%%%%%%%%%%%%%%%%%%%%%%%%%%%%%%%%%%%%%%%%%%%%%%%%%%%%%%%%%
\section{Self-Gravitational Corrections to Cardy-Verlind Formula}
We are interested primarily in the corrections to the entropy (\ref{entro}) that arise in the context of KKW
analysis \cite{KKW} \footnote{Further developments in KKW analysis can be found in \cite{corrections}.} due to
self-gravitational effect. Here we adopt the analysis as presented by Medved \cite{medved1} and thus we consider
the incoming radiation from the cosmological horizon and, for the moment, ignore the outgoing radiation from the
black hole horizon. Therefore, a pair of particles is spontaneously created just outside the cosmological horizon.
The positive-energy particle is viewed as an incoming particle since it tunnels through the cosmological horizon
and arrives to the bulk space. Thus, the black hole mass is increased while the background energy is lowered since
the negative-energy particle remains behind the cosmological horizon. Let us remind that the key point to the KKW
analysis is that the total energy of the spacetime under study is kept fixed while the black hole mass is allowed
to vary. We therefore expect a black hole of initial mass $M$ to have a final mass of $M+\omega$ where $\omega$ is
the energy of the emitted particle. The point of particle creation, $r_{i}$, is the radius of the cosmological
horizon, i.e. $r_c$,
\bea
r_{i}^{2}=\frac{l^{2}}{2}\left(1+\sqrt{1-\frac{4\varepsilon_3 M}{l^{2}}}\right)
\label{radi3}
\eea
while the classical turning point of motion is given by
\bea
r_{f}^{2}=\frac{l^{2}}{2}\left(1+\sqrt{1-\frac{4\varepsilon_3
(M+\omega)}{l^{2}}}\right)
\hspace{1ex}.
\label{radi4}
\eea
To first order in the emitted energy ($\omega$), the
aforementioned radii are related as follows \be r^{2}_{f}=r^{2}_{i}\left(1-\omega \frac{\varepsilon_3}{r^{2}_{i}
\sqrt{1-\frac{4\varepsilon_3 M}{l^{2}}}}\right) \ee Consequently, the change in the entropy of the $SdS_5$ black
hole during the process of emission takes the form \bea \Delta
S=S_f-S_i=\frac{V_3}{4G_5}(r_{f}^{3}-r_{i}^{3})=-\frac{\omega}{T(\omega)} \label{chang} \eea where $S_f$ is the
modified entropy of the $SdS_5$ black hole due to the self-gravitational effect, $S_i$ is the standard formula for
the entropy (Bekenstein-Hawking entropy) derived when the black hole mass is kept fixed, and $T(\omega)$ is the
corrected temperature of the cosmological horizon due to the self-gravitational effect.
The expression for the modified temperature of the cosmological horizon is given as
\be
T(\omega)=\frac{\left(2r_{i}^{2}-l^{2}\right)}{2\pi l^{2}r_{i}}+ \mathcal{O}(\omega)
\label{modtemp}
\ee
It is obvious that, to first order in $\omega$, the modified temperature is the Hawking temperature of the cosmological
horizon, i.e. $T_{c}$. Substituting equation (\ref{modtemp}) into equation (\ref{chang}), we evaluate the corrected
entropy of the  $SdS_5$ black hole due to the self-gravitational effect \be S_{f}=S_{i}-\frac{2\pi
l^{2}r_{i}}{\left(2r_{i}^{2}-l^{2}\right)}\omega + \mathcal{O}(\omega^2) \hspace{1ex}. \label{modent} \ee Since we
are primarily interested to keep the leading-order correction (that is first order in $\omega$), we drop the last
term in equation (\ref{modent}) and the corrected entropy of the $SdS_5$ black hole reads \be
S_{f}=S_{i}-\frac{2\pi l^{2}r_{i}}{\left(2r_{i}^{2}-l^{2}\right)}\omega \hspace{1ex}. \label{modentropy} \ee

\par\noindent
It is interesting that the semiclassical correction to the black hole entropy is negative. This may be compared
with the possibility for SdS black hole entropies to be negative in the context of higher derivative gravitational
theory \cite{cvetic}. In this work, the negativity of black hole entropy was regarded as an indication of a new
type of instability in asymptotically de Sitter black hole physics. Thus, the self-gravitational corrections
presented here may also try to put the $SdS_5$ black hole under consideration to a less stable state.

\par\noindent
Due to the dS/CFT correspondence, we are now ready to evaluate the corrections to the Cardy-Verlinde formula for
the entropy of the $SdS_5$ black hole. An important physical quantity that gets involved in this calculation is
the Casimir energy which is defined as the violation of the Euler identity and for the spacetime under study is
given by \bea E_C=4E_{4}-3T_{c}S_i \label{cas} \eea where the four-dimensional energy $E_4$ which can be derived
from the FRW equation of motion for a brane propagating in the $SdS_{5}$ bulk space, is given by \cite{verlinde}
\bea E_4=\pm\frac{l}{r}M=\frac{l}{r}E \label{ener2} \eea where $`+$' corresponds to the black hole horizon and $`-$'
to the cosmological horizon. The temperature associated with the $4$-dimensional brane, $T^{brane}_{c}$,
should differ from expressions (\ref{costem}) and (\ref{blatem}) by a similar factor \cite{verlinde} \bea
T^{brane}_{c}=\frac{l}{r}T_{c} \label{costem3} \eea It is easily seen that, to first order in $\omega$, the
temperature of the brane is unchanged since the temperature of the cosmological horizon, i.e. $T_{c}$, is unchanged as
stated before. The modified Casimir energy  is given by
\be
\mathcal{E}_C=4\mathcal{E}_{4}-3\mathcal{T}^{brane}_{c}\mathcal{S}_{i}
\label{modcas}
\ee
where calligraphic letters denote that the corresponding quantity
is modified due to the self-gravitational effect. The modified
four-dimensional energy is
\be
\mathcal{E}_4=E_{4}-\frac{l}{r}\omega
 \label{modener2}
 \ee
while the last term in (\ref{modcas}) is given by
\bea
\mathcal{T}^{brane}_{c}\mathcal{S}_{i} &=& T^{brane}_{c}S_{f}\nn\\
&=& T^{brane}_{c}S_{i}-\frac{T^{brane}_{c}}{T_{c}}\omega\nn\\
&=& T^{brane}_{c}S_{i}-\frac{l}{r}\omega
\label{modTS}
\eea
Substituting equations (\ref{modener2}) and
(\ref{modTS}) in (\ref{modcas}), the modified Casimir energy takes the form
\be
\mathcal{E}_{C}=E_{C}-\frac{l}{r}\omega \hspace{1ex}. \label{modCAS} \ee
Due to the self-gravitational
corrections, the modified Cardy-Verlinde formula for the entropy of the $SdS_{5}$ black hole is given as \be
\mathcal{S}_{CFT}=\frac{2\pi r}{3}\sqrt{{\Big|}\mathcal{E}_C(2\mathcal{E}_4-\mathcal{E}_{C})\Big|} \hspace{1ex}
\label{cv}
\ee
Using equations (\ref{modener2}) and (\ref{modCAS}), the modified Cardy-Verlinde entropy formula becomes
\be \mathcal{S}_{CFT}=\frac{2\pi r}{3}\sqrt{\Bigg| \left[E_{C}-\frac{l}{r}\omega\right]
\left[\left(2E_{4}-E_{C}\right)-\frac{l}{r}\omega\right]\Bigg|} \hspace{1ex} \ee and keeping terms up to first
order in the emitted energy $\omega$, it takes the form

\be \mathcal{S}_{CFT}=S_{CFT}\left(1-\varepsilon \omega\right) \hspace{1ex} \label{modS} \ee where the small
parameter $\varepsilon$ is given by \be \varepsilon=\frac{l}{r}\frac{E_{4}}{E_{C}\left(2E_{4}-E_{C}\right)}
\hspace{1ex}. \label{epsilon} \ee A welcomed but not unexpected result is that there is no entropy bound violation
due to self-gravitational corrections to the Cardy-Verlinde entropy\footnote{A violation of the holographic
entropy bound was observed when self-gravitational corrections to the Cardy-Verlinde entropy formula of the
two-dimensional Ach\'ucarro-Ortiz black hole were included \cite{setelias2} (see also \cite{mignemi}).}.
%%%%%%%%%%%%%%%%%%%%%%%%%%%%%%%%%%%%%%%%%%%%%%%%%%%%%%%%%%%%%%%%%%%%%%%%%%%%%%%%%%%%%%%%%%%%%%%%
%%%%%%%%%%%%%%%%%%%%%%%%%%%%%%%%%%%%%%%%%%%%%%%%%%%%%%%%%%%%%%%%%%%%%%%%%%%%%%%%%%%%%%%%%%%%%%%%
%%%%%%%%%%%%%%%%%%%%%%%%%%%%%%%%%%%%%%%%%%%%%%%%%%%%%%%%%%%%%%%%%%%%%%%%%%%%%%%%%%%%%%%%%%%%%%%%
%%%%%%%%%%%%%%%%%%%%%%%%%%%%%%%%%%%%%%%%%%%%%%%%%%%%%%%%%%%%%%%%%%%%%%%%%%%%%%%%%%%%%%%%%%%%%%%%
\section{Self-Gravitational Corrections to FRW Brane Cosmology}
 We now consider a $4$-dimensional brane in the
$SdS_5$ black hole background. This $4$-dimensional brane can be regarded as the boundary of the
$SdS_{5}$ bulk background. Let us first replace the radial coordinate $r$ with $a$ and hence
the line element described by equation (\ref{metric1}) now takes the form
\bea
ds^{2}=-h(a)dt^{2}+\frac{1}{h(a)}da^{2}+a^{2}d\Omega_{(3)}^{2}
\hspace{1ex},
\label{met}
\eea
where
\be
h(a)=\left(1-\frac{a^{2}}{l^{2}}-\frac{\mu}{a^{2}}\right)
\label{metric}
\ee
and
\be \mu=M\varepsilon_{3} \ee is the black
hole mass parameter. It was shown that by reduction from the $SdS_5$ background (\ref{met}) and by imposing the
condition \be -h(a)\left(\frac{\partial t}{\partial \tau}\right)^{2}+ \frac{1}{h(a)}\left(\frac{\partial
a}{\partial \tau}\right)^{2}=-1 \label{condition2} \ee where $\tau$ is a new time parameter, one obtains an FRW
metric for the $4$-dimensional timelike brane \be ds_{(4)}^{2}=-d\tau^{2}+a^{2}(\tau)d\Omega_{(3)}^{2}
\hspace{1ex}. \ee Thus, the $4-$dimensional FRW equation describes the motion of the brane universe in the $SdS_5$
background. It is easy to see that the matter on the brane can be regarded as radiation and consequently, the
field theory on the brane should be a CFT.

\par\noindent
Within the context of AdS/CFT correspondence, Savonije and Verlinde studied the CFT/FRW-cosmology
relation from the Randall-Sundrum type braneworld perspective \cite{savonije}. They showed that the entropy
formulas of the CFT coincide with the Friedmann equations when the brane crosses the black hole horizon.
Analogously, one can assume holographic relations between the FRW universe and the boundary CFT which is dual to
the $SdS_5$ bulk background in accordance with dS/CFT correspondence. In the case of a $4$-dimensional
spacelike\footnote{In this work we are interested in the radiation coming in the bulk space through the
cosmological horizon. Timelike brane, i.e a brane that has a Minkowskian metric, can only cross the black hole
horizon. On the contrary, a spacelike brane, i.e. a brane with Euclidean metric, is able to cross both the black
hole horizon and the cosmological horizon.} brane\footnote{In order to derive the $4$-dimensional spacelike brane,
the imposed condition (\ref{condition2}) has to be slightly changed by replacing the $`-$' with a $`+$' on the
right-hand side of it. } \be ds_{(4)}^{2}=d\tau^{2}+a^{2}(\tau)d\Omega_{(3)}^{2} \hspace{1ex}, \ee one of the
identifications that supports the CFT/FRW-cosmology relation  is \be
H^{2}=\left(\frac{2G_4}{V}\right)^{2}\mathcal{S}^{2} \label{huble} \ee where $H$ is the Hubble parameter defined
by $H=\frac{1}{a}\frac{da}{d\tau}$ and V is the volume of the $3$-sphere ($V=a^{3}V_{3}$). The $4$-dimensional
Newton constant $G_4$ is related to the $5-$dimensional one $G_5$ by \be
 G_{4}=\frac{2}{l}G_{5}
 \hspace{1ex}.
\ee It was shown that at the moment that the $4$-dimensional spacelike brane crosses the cosmological horizon,
i.e. when $a=a_{c}$, the CFT entropy and the entropy of the $SdS_5$ black hole are identical. By substituting
(\ref{modS}) into (\ref{huble}), we obtain the self-gravitational corrections to the motion of the CFT-dominated
brane \be H^{2}=\left(\frac{2G_4}{V}\right)^{2}S_{CFT}^{2}\left(1-\varepsilon \omega\right)^{2} \label{modH}
\hspace{1ex}. \ee It is obvious that from the first term on the right-hand side of (\ref{modH}) we get the
standard Friedmann equation with the appropriate normalization \be H^{2}=\frac{1}{a_{c}^{2}}-\frac{8\pi
G_{4}}{3}\rho \ee where $\rho$ is the energy density defined by $\rho=|E_{4}|/V$. Therefore, the correction to the
FRW equation due to the self-gravitation effect is expressed by the second term in the right-hand side of equation
(\ref{modH}). Keeping terms up to first order in the emitted energy ($\omega$), the modified Hubble equation due
to the self-gravitation corrections is \be H^{2}=\frac{1}{a_{c}^{2}}-\frac{8\pi G_{4}}{3}\rho
-2E_{4}\frac{l}{a_{c}} \left(\frac{2G_{4}}{V}\right)^{2}\left(\frac{2\pi a_{c}}{3}\right)^{2}\omega \hspace{1ex}.
\ee Taking into account that all quantities should be evaluated on the cosmological horizon, the modified Hubble
equation, i.e. the first Friedmann equation, takes the form \be H^{2}=\frac{1}{a_{c}^{2}}-\frac{8\pi G_{4}}{3}\rho
+ \frac{8\pi G_{4}}{3}\left[\frac{4\pi G_{4}}{3}\frac{l}{a_{c}^{2}V_{3}}\rho\right]\omega \label{modHubble} \ee
where the volume $V$ is given by $a_{c}^{3}V_{3}$. At this point it should be stresses that our analysis was up to
now restricted to the spatially flat ($k=+1$) spacelike brane.
\par
We will now extend the aforesaid analysis. We therefore consider an arbitrary scale factor $a$ and include a
general $k$ taking values ${+1,0,-1}$ in order to describe, respectively, the elliptic, flat, and hyperbolic
horizon geometry of the $SdS_5$ bulk black hole. The modified Hubble equation is now given by \be
H^{2}=\frac{k}{a^{2}}-\frac{8\pi G_{4}}{3}\rho + \frac{8\pi G_{4}}{3}\left[\frac{4\pi
G_{4}}{3}\frac{l}{a^{2}V_{3}}\rho\right]\omega \label{gmodHubble} \ee where the volume $V$ is now given by
$a^{3}V_{3}$   since all quantities that appear in equation (\ref{gmodHubble}) are defined for an arbitrary scale
factor $a$.

The first term in the right-hand side of equation (\ref{gmodHubble}) represents the curvature contribution to the
brane motion. The second term can be regarded as the contribution from the radiation and  it redshifts as
$a^{-4}$. The last term in the right-hand side of equation (\ref{gmodHubble}) is the self-gravitational correction
to the motion of $4$-dimensional spacelike brane moving in the $5$-dimensional Schwarzschild-de Sitter bulk
background. Since this term goes like $a^{-6}$, it is obvious that it is dominant at early times of the brane
evolution while at late times the second term, i.e. the radiative matter term, dominates and thus the last term
can be neglected. There are three different ways to interpret the last term in the right-hand side of equation
(\ref{gmodHubble}).
\par\noindent
The first choice is that the last term in right-hand side of equation (\ref{gmodHubble}) can be regarded as a
$\rho^{2}$ term. Due to the opposite sign with respect to the $\rho$ term, the FRW universe under consideration is
led to an inevitable bounce \cite{bounce}. However, in order this scenario to be materialized, one has to treat the
emitted energy $\omega$ as a $4$-dimensional quantity. Thus, it has to be scaled according to
\be
\omega=\frac{l}{a}\omega_{5}
\ee
where $\omega_{5}$ has to be the shell of energy emitted by the $5$-dimensional Schwarzschild-de Sitter bulk
background as discussed in the preceding section within the context of KKW analysis.

\par\noindent
The second choice is that the last term can be regarded as an anisotropy energy density. In this case, the
modified Hubble equation (\ref{gmodHubble}) takes the form \be H^{2}=\frac{k}{a^{2}}-\frac{8\pi G_{4}}{3}\rho
+\frac{8\pi G_{4}}{3}\rho_{shear} \label{shearHubble} \ee where the anisotropy energy density is given by \be
\rho_{shear}=\frac{\Sigma^{2}}{a^{6}} \label{shear} \ee
with
\be
\Sigma^{2}=\frac{4\pi
G_{4}}{3}\left(\frac{l}{V_{3}}\right)^{2}M\omega
\hspace{1ex}.
\ee
It should be mentioned that in standard cosmology where there are no corrections,
the shear anisotropy term is a product of the anisotropic character of
the brane geometry \cite{shear}.

Finally, the third choice which looks more appealing, is  to regard the last term in the right-hand side of equation
(\ref{gmodHubble}) as stiff matter \cite{stiff1}. In particular, within the context of dS/CFT correspondence,
Medved  considered\footnote{It should be noted that Medved considered a RN-dS backround spacetime of arbitrary
dimensionality. Here, for simplicity reasons, we reproduce his results for the case of $n=3$, i.e. the RN-dS bulk
background is five-dimensional.} a Reissner-Nordstrom-de Sitter bulk background \cite{stiff2} \be
ds^{2}=-h(a)dt^{2}+\frac{1}{h(a)}da^{2}+a^{2}d\Omega_{(3)}^{2}, \ee where \be
h(a)=1-\frac{a^{2}}{l^{2}}-\frac{\varepsilon_{3}M}{a^{2}}+\frac{3\varepsilon_{3}^{2}Q^{2}}{16a^{4}} \hspace{1ex}.
\ee and $M$, $Q$ represent the black hole mass and charge, respectively. It was showed that the brane evolution is
described by a Friedmann-like equation for radiative matter along with a stiff-matter contribution \bea
H^{2}&=&\frac{1}{a^{2}}-\frac{\varepsilon_{3}M}{a^{4}}+
\frac{3\varepsilon_{3}^{2}Q^{2}}{16a^{6}}\nn\\
&=&\frac{1}{a^{2}}-\frac{8\pi G_{4}}{3}\rho+ \frac{3\varepsilon_{3}^{2}Q^{2}}{16a^{6}} \label{stiff} \hspace{1ex}.
\eea
If the following condition is satisfied
\be
Q^{2}=\frac{8}{3}M\omega
\hspace{1ex},
\ee
then equations (\ref{gmodHubble})  and (\ref{stiff})
are identical (for the case $k=+1$) and the last term in the right-hand side of
equation (\ref{gmodHubble}) is then regarded as stiff matter.

\par\noindent
In passing, one should not worry about the opposite
(wrong) sign of the stiff-matter term with respect to the radiative matter term. The reason is that the
$4$-dimensional spacelike brane has an Euclidean metric and typically such rotations (from the Minkowskian to the
Euclidean sector of a theory) are followed by the transformation $Q^{2}\rightarrow -Q^{2}$ \cite{gibbons}.

\par
Finally a couple of comments are in order.
Firstly, if one takes the $\tau$ derivative of the modified Hubble
equation (\ref{gmodHubble}) then the second modified Friedmann equation is obtained \be
\dot{H}=-\frac{k}{a^{2}}+\frac{16\pi G_{4}}{3}\rho - 8\pi G_{4}\left[\frac{4\pi
G_{4}}{3}\frac{l}{a^{2}V_{3}}\rho\right]\omega \hspace{1ex}. \ee
Secondly, one can rewrite the modified Hubble
equation (\ref{gmodHubble}) in a such a way that it represents the conservation of energy of a point particle
moving in a one-dimensional effective potential, $V(a)$, \be \left(\frac{da}{d\tau}\right)^{2}=k-V(a)
\label{pointparticle} \ee where the variable $a$ represents the position of the particle and the modified
effective potential due to the self-gravitational effect reads \bea V(a)&=&\frac{8\pi G_{4}}{3}a^{2}\rho -
\frac{8\pi G_{4}}{3}\left(\frac{4\pi l G_{4}}{3V_{3}}\rho\right)\omega\nn\\
&=&\frac{\mu}{a^{2}}-\frac{\beta}{a^{4}} \label{pot} \eea where \be \beta=2\left(\frac{4\pi
G_{4}l}{3V_{3}}\right)^{2}M\omega \hspace{1ex}. \label{beta} \ee The first term in the expression for the
effective potential (\ref{pot}) is called ``dark radiation'' term.

\par\noindent
Let us now recall the one-dimensional effective potential that appears in (\ref{pointparticle}) for the case of
standard FRW cosmology. In this framework, there are no corrections and hence the first term in (\ref{pot}) is
actually the only term in the expression for the effective potential \be V(a)=\frac{\mu}{a^{2}} \hspace{1ex}. \ee
The $4$-dimensional spacelike FRW brane exists in the regions where $V(a)\leq k$ so that $H^{2}\geq 0$. It is
clear that the only viable scenario is the case of $k=+1$ which is the spherical brane (spatially closed universe).
The spherical brane starts at $a=+\infty$ and reaches its minimum size at $a_{min}=\sqrt{\mu}$ and then
re-expands.

\par\noindent
On the contrary, when one includes quantum corrections, i.e. logarithmic corrections, then the effective potential
takes the form \cite{logcor} \be V(a)=\frac{\mu}{a^{2}}+3\frac{\gamma}{a}\log \left(\frac{a}{\gamma l^{2}}\right)
\label{log} \ee where $\gamma$ is given by \be \gamma=\frac{2G_{4}}{V_{3}l} \hspace{1ex}. \ee It is evident that
in this case the behavior of the effective potential depends on the parameters $\mu$ and $\gamma$ (which in turn
depends on $G_{4}$, $V_{3}$ and $l$). If one assumes that $\gamma\leq \sqrt{\mu}$, then the behavior of the
effective potential is almost the same as described above for the case of standard FRW cosmology. But if $\gamma$
is large compared to $\sqrt{\mu}$, then the behavior of the effective potential is modified\footnote{For a
complete analysis on the behavior of the effective potential when logarithmic corrections are included see
\cite{logcor}.}.

\par\noindent
Comparing the expression for the effective potential when self-gravitational corrections (\ref{pot}) are taken
into account with the one when the logarithmic corrections (\ref{log}) are present, the important difference is
that the self-gravitational corrections are subtractive while the logarithmic ones are additive (with respect to
the ``dark radiation'' term). Furthermore, by substituting equation (\ref{pot}) into (\ref{pointparticle}), we
obtain
\be
\left(\frac{da}{d\tau}\right)^{2}=k-\frac{\mu}{a^{2}}+\frac{\beta}{a^{4}}
\label{pointparticle2}
\hspace{1ex}.
\ee
It is remarkable that the third term which is the contribution due to the the self-gravitational
effect is so small for large values of $a$, i.e. at late times of the brane evolution, that it can be neglected.
In this case the behavior of the effective potential is qualitative similar to that of the standard FRW cosmology
presented before. On the contrary, for small values of $a$, the contribution of the self-gravitational corrections
is very important. Thus, the third term dominates with respect to the second term, i.e. the ``dark radiation''
term. The asymptotic behavior of the brane is as follows.
\begin{itemize}
\item {\bf k=-1}\\
The $4$-dimensional brane expands from an initial state of vanishing spatial  volume, $a=0$, and reaches its
maximal size $a_{max}$ when the constraint equation $V(a)=-1$ is satisfied. Then the negatively curved brane
undergoes a recollapse to zero volume.
\item {\bf k=0}\\
As in the case of negatively curved brane, the $4$-dimensional flat brane starts form $a=0$ and reaches its
maximal size $a_{max}$ when the constraint equation $V(a)=0$ is satisfied. A recollapse to zero volume follows.
\item {\bf k=+1}\\
In this case there are a number of different outcomes for the positively curved brane. If the constraint equation
$V(a)<+1$ is always satisfied, then the $4$-dimensional positively curved brane expands to infinity.  But if the
potential exceeds the critical value of unity then the brane reverse its direction of motion. Solving the equation
$V(a)=+1$, two solutions are obtained $a_{1}$ and $a_{2}$ (say $a_{1}< a_{2}$). Therefore, if the constraint
equation $V(a)<+1$ is satisfied, then the $4$-dimensional positively curved brane either starts from $a=0$,
reaches its maximal size $a_{1}$ and finally undergoes a recollapse to zero volume, or starts from infinite
spatial volume, undergoes a collapse to a minimal size $a_{2}$ and reexpands to infinity.
This is the behavior of a ``bouncing'' universe.
\end{itemize}
%%%%%%%%%%%%%%%%%%%%%%%%%%%%%%%%%%%%%%%%%%%%%%%%%%%%%%%%%%%%%%%%%%%%%%%%%%%%%%%%%
%%%%%%%%%%%%%%%%%%%%%%%%%%%%%%%%%%%%%%%%%%%%%%%%%%%%%%%%%%%%%%%%%%%%%%%%%%%%%%%%%
%%%%%%%%%%%%%%%%%%%%%%%%%%%%%%%%%%%%%%%%%%%%%%%%%%%%%%%%%%%%%%%%%%%%%%%%%%%%%%%%%
%%%%%%%%%%%%%%%%%%%%%%%%%%%%%%%%%%%%%%%%%%%%%%%%%%%%%%%%%%%%%%%%%%%%%%%%%%%%%%%%%
\section{Conclusions}
One of the striking results for the dynamic dS/CFT correspondence is that the Cardy-Verlinde formula on the
CFT-side coincides with the first Friedmann equation (Hubble equation) in cosmology when the brane crosses the
horizon, i.e. when $a=a_{c}$, of the $SdS_5$ black hole. This means that the Hubble equation knows the thermodynamics of the
CFT. In this paper we have considered the dynamics of a $4$-dimensional spacelike FRW brane propagating in an
$5$-dimensional dS  bulk space containing  a Schwarzschild black hole. Taking into account the semiclassical
corrections to the black hole entropy that arise as a result of the self-gravitational effect,  and employing the
dS/CFT correspondence, we obtained the self-gravitational corrections to the Cardy-Verlinde formula. A welcomed
but not unexpected result was that the modified entropy doesn't violate any entropy bound since the additional
term due to the self-gravitation effect is subtractive. These self-gravitational corrections to the Cardy-Verlinde
entropy formula express the existence of a deep connection between semiclassical thermodynamics and de Sitter
holography.

\par\noindent
Furthermore, the self-gravitational corrections to the associated Friedmann-like brane equations are obtained. The
additional term in the Hubble equation due to the self-gravitation effect goes as $a^{-6}$. We presented several
ways of interpreting this term. It seems that the most appealing choice since it doesn't require a scaling of the
emitted energy, or a generalization of the FRW brane geometry to an anisotropic one, is the stiff matter. Thus,
the self-gravitational corrections act as a source of stiff matter contrary to standard FRW cosmology where the
charge of the black hole plays this role.

\par\noindent
Finally, we rewrite the Hubble equation in a way that represents the conservation equation of energy of a point
particle moving in a one-dimensional effective potential. The self-gravitational corrections to the effective
potential and, consequently, to the conservation equation of energy of the point particle are presented. The term that
describes the semiclassical corrections becomes dominant at early times of the brane evolution. On the
contrary, at late times the ``dark radiation'' term is the dominant one. A short analysis of the asymptotic
behavior of the brane follows. It is remarkable that in the case of the positively curved brane, there is the
possibility of getting a bouncing universe. A feature that is not possible in the framework of standard FRW
cosmology.

\par\noindent
At this point a couple of questions are raised. First, is it possible for an observer that lives on the brane and
observes the stiff matter to discriminate the kind of source that produces the stiff matter? The second question
arises when one includes both semiclassical and quantum corrections. Which is the dominant correction and when
does this dominance take place during the brane evolution?
We plan to address these interesting issues in a future work.
%%%%%%%%%%%%%%%%%%%%%%%%%%%%%%%%%%%%%%%%%%%%%%%%%%%%%%%%%%%%%%%%%%%%%%%%%%%%%%%%%
%%%%%%%%%%%%%%%%%%%%%%%%%%%%%%%%%%%%%%%%%%%%%%%%%%%%%%%%%%%%%%%%%%%%%%%%%%%%%%%%%
%%%%%%%%%%%%%%%%%%%%%%%%%%%%%%%%%%%%%%%%%%%%%%%%%%%%%%%%%%%%%%%%%%%%%%%%%%%%%%%%%
%%%%%%%%%%%%%%%%%%%%%%%%%%%%%%%%%%%%%%%%%%%%%%%%%%%%%%%%%%%%%%%%%%%%%%%%%%%%%%%%%
%%%%%%%%%%%%%%%%%%%%%%%%%%%%%%%%%%%%%%%%%%%%%%%%%%%%%%%%%%%%%%%%%%%%%%%%%%%%%%%%%
\section{Acknowledgments}
E.C.V. has been supported by the European Research and Training Network ``EUROGRID-Discrete Random Geometries: from Solid
State Physics to Quantum Gravity" (HPRN-CT-1999-00161).

%%%%%%%%%%%%%%%%%%%%%%%%%%%%%%%%%%%%%%%%%%%%%%%%%%%%%%%%%%%%%%%%%%%%%%%%%%%%%%%%%
%%%%%%%%%%%%%%%%%%%%%%%%%%%%%%%%%%%%%%%%%%%%%%%%%%%%%%%%%%%%%%%%%%%%%%%%%%%%%%%%%
%%%%%%%%%%%%%%%%%%%%%%%%%%%%%%%%%%%%%%%%%%%%%%%%%%%%%%%%%%%%%%%%%%%%%%%%%%%%%%%%%
%%%%%%%%%%%%%%%%%%%%%%%%%%%% Bibliography %%%%%%%%%%%%%%%%%%%%%%%%%%%%%%%%%%%%%%%

%%%%%%%%%%%%%%%%%%%%%%%%%%%%%%%%%%%%%%%%%%%%%%%%%%%%%%%%%%%%%%%%%%%%%%%%%%%%%%%%%%%%%%%%%%%%%%%%%%
%%%%%%%%%%%%%%%%%%%%%%%%%%%%%%%%%%%%%%%%%%%%%%%%%%%%%%%%%%%%%%%%%%%%%%%%%%%%%%%%%%%%%%%%%%%%%%%%%%

\begin{thebibliography}{99}
%%%%%%%%%%%%%%%%%%%%%%%%%%%%

\bibitem{hawking1}
S. W. Hawking, Commun. Math. Phys. {\bf43} (1975) 199.

%%%%%%%%%%%%%%%%%%%%%%%%%%%%

\bibitem{KKW}
E. Keski-Vakkuri and P. Kraus, Phys. Rev. D {\bf54} (1996) 7407, hep-th/9604151;
P. Kraus and F. Wilczek, Nucl. Phys. B {\bf433} (1995) 403, gr-qc/9408003;
P. Kraus and F. Wilczek, Nucl. Phys. B {\bf437} (1995) 231, hep-th/9411219;
E. Keski-Vakkuri and P. Kraus, Nucl. Phys. B {\bf491} (1997) 249, hep-th/9610045;
M. K. Parikh and F. Wilczek, Phys. Rev. Lett. {\bf85} (2000) 5042, hep-th/9907001.

\bibitem{corrections}
Y. Kwon, Il Nuovo Cimento B {\bf115} (2000) 469;
S. Hemming and E. Keski-Vakkuri, Phys. Rev. D {\bf64} (2001) 044006, gr-qc/0005115;
E. C. Vagenas, Phys. Lett. B {\bf503} (2001) 399, hep-th/0012134;
E. C. Vagenas, Mod. Phys. Lett. A {\bf17}  (2002) 609, hep-th/0108147;
E. C. Vagenas, Phys. Lett. B {\bf533} (2002) 302, hep-th/0109108;
A. J. M. Medved, Class. Quant. Grav. {\bf19} (2002) 589, hep-th/0110289;
E. C. Vagenas, Phys. Lett. B {\bf559} (2003) 65, hep-th/0209185;
M. K. Parikh, hep-th/0402166.


%%%%%%%%%%%%%%%%%%%%%%%%%%%%%

\bibitem{Ries98} A. G. Riess et al., Astron. J. {\bf116} (1998) 1009, astro-ph/9805201;
S. Perlmutter et al., Astrophys. J. {\bf517} (1999) 565, astro-ph/9812133;
P. de Bernardis et al.,  Nature {\bf404} (2000) 955, astro-ph/0004404;
C. L. Bennett et al., Astrophys. J. Suppl. {\bf148} (2003) 1, astro-ph/0302207;
M. Tegmark et al., Phys. Rev. D {\bf69} (2004) 103501, astro-ph/0310723.

%%%%%%%%%%%%%%%%%%%%%%%%%%%%%

\bibitem{ds/cft}
A. Strominger, JHEP {\bf0110} (2001) 034, hep-th/0106113;
A. Strominger, JHEP {\bf0111} (2001) 049, hep-th/0110087;
For review, see M. Spradlin, A. Strominger and A. Volovich, hep-th/0110007;
M. -I. Park, Phys. Lett. B {\bf440} (1998) 275, hep-th/9806119;
M. -I. Park, Nucl. Phys. B {\bf544} (1999) 377, hep-th/9811033;
S. Nojiri and S. D. Odintsov, Phys. Lett. B {\bf519} (2001) 145, hep-th/0106191;
S. Nojiri and S. D. Odintsov, JHEP {\bf0112} (2001) 033, hep-th/0107134;
S. Nojiri, S. D. Odintsov and S. Ogushi, Phys. Rev. D {\bf65} (2002) 023521, hep-th/0108172;
S. Nojiri and S. D. Odintsov, Phys. Lett. B {\bf523} (2001) 165, hep-th/0110064;
D. Klemm, Nucl. Phys. B {\bf625} (2002) 295, hep-th/0106247;
J. Bros, H. Epstein and U. Moschella, Phys. Rev. D {\bf65} (2002) 084012, hep-th/0107091;
T. Shiromizu, D. Ida and T. Torii, JHEP {\bf0111} (2001) 010, hep-th/0109057;
C. M. Hull, JHEP {\bf0111} (2001) 012, hep-th/0109213;
S. Cacciatori and D. Klemm, Class. Quant. Grav. {\bf19} (2002) 579, hep-th/0110031;
B. McInnes, Nucl. Phys. B {\bf627} (2002) 311, hep-th/0110062;
Y. S. Myung, Mod. Phys. Lett. A {\bf16} (2001) 2353, hep-th/0110123;
A. C. Petkou and G. Siopsis, JHEP {\bf0202} (2002) 045, hep-th/0111085;
R. G. Cai, Phys. Lett. B {\bf525} (2002) 331, hep-th/0111093;
A. Padilla, Phys. Lett. B {\bf528} (2002) 274, hep-th/0111247;
E. Halyo, JHEP {\bf0203} (2002) 009, hep-th/0112093;
Y. S. Myung, Phys. Lett. B {\bf531} (2002) 1, hep-th/0112140;
R. G. Cai, Nucl. Phys. B {\bf628} (2002) 375, hep-th/0112253;
U. H. Danielsson, JHEP {\bf0203} (2002) 020, hep-th/0110265;
S. Ogushi, Mod. Phys. Lett. A {\bf17} (2002) 51, hep-th/0111008;
J. P. Gregory and A. Padilla, Class. Quant. Grav. {\bf19} (2002) 4071, hep-th/0204218;
S. Ogushi, hep-th/0210055;
M. R. Setare, Mod. Phys. Lett. A {\bf17} (2002) 2089, hep-th/0210187;
M. R. Setare and R. Mansouri, Int. J. Mod. Phys.  A {\bf18} (2003) 4443, hep-th/0210252;
M. R. Setare and M.B. Altaie, Eur. Phys. J. C {\bf30} (2003) 273,  hep-th/0304072;
J. P. Gregory and A. Padilla, Class. Quant. Grav. {\bf20} (2003) 4221, hep-th/0304250;
D.~Astefanesei, R.~Mann and E.~Radu, JHEP {\bf0401} (2004) 029, hep-th/0310273;
M. R. Setare, hep-th/0405010.

%%%%%%%%%%%%%%%%%%%%%%%%%%%%

\bibitem{odi}
S. Nojiri, S. D. Odintsov and S. Ogushi, Int. J. Mod. Phys. A {\bf17} (2002) 4809, hep-th/0205187.

%%%%%%%%%%%%%%%%%%%%%%%%%%%%

\bibitem{verlinde} E. Verlinde, hep-th/0008140.

%%%%%%%%%%%%%%%%%%%%%%%%%%%%%

%%%%%%%%%%%%%%%%%%%%%%%%%%%%%

\bibitem{logcorrections1}
R. K.  Kaul and P. Majumdar, Phys. Rev. Lett. {\bf 84} (2000) 5255, gr-qc/0002040;
S. Das, P. Majumdar and R. K. Bhaduri, Class. Quant. Grav. {\bf19} (2002) 2355, hep-th/0111001;
D. V. Fursaev, Phys. Rev. D {\bf51} (1995) 5352, hep-th/9412161;
R. B. Mann and S. N. Solodukhin, Nucl. Phys. B {\bf523} (1998) 293, hep-th/9709064;
A. J. M. Medved and G. Kunstatter, Phys. Rev. D {\bf60} (1999) 104029, hep-th/9904070;
O. Obregon, M. Sabido and V. I. Tkach, Gen. Rel. Grav. {\bf33} (2001) 913, gr-qc/0003023;
S. Das, R. K. Kaul and P. Majumdar, Phys. Rev. D {\bf63} (2001) 044019, hep-th/0006211;
A. J. M. Medved and G. Kunstatter, Phys. Rev. D {\bf63} (2001) 104005, hep-th/0009050;
T. R. Govindarajan, R. K. Kaul and V. Suneeta, Class. Quant. Grav. {\bf18} (2001) 2877, gr-qc/0104010;
S. Mukherji and S. S. Pal, JHEP {\bf0205} (2002) 026, hep-th/0205164;
S. Das, hep-th/0207072;
J. E. Lidsey, S. Nojiri, S. D. Odintsov and S. Ogushi, Phys. Lett. B {\bf544} (2002) 337, hep-th/0207009;
G. Gour, Phys. Rev. D {\bf66} (2002) 104022, gr-qc/0210024;
A. Chatterjee and P. Majumdar, gr-qc/0303030;
M. M. Akbar and S. Das, Class.Quant.Grav. {\bf21} (2004) 1383, hep-th/0304076;
M. R. Setare, Eur. Phys. J. C {\bf33}  (2004) 555, hep-th/0309134;
M. R. Setare,  Phys. Lett. B {\bf573} (2003) 173, hep-th/0308106;
M. R. Setare,  hep-th/0311106;
S. Das, hep-th/0403202.
%%%%%%%%%%%%%%%%%%%%%%%%%%%%%



\bibitem{carlip1}
S. Carlip, Class. Quant. Grav. {\bf17} (2000) 4175, gr-qc/0005017.
\bibitem{carlip2}
S. Carlip, Class. Quant. Grav. {\bf15} (1998) 3609, hep-th/9806026.

%%%%%%%%%%%%%%%%%%%%%%%%%%%%%%%%%

\bibitem{setelias1}
M. R. Setare and E. C. Vagenas, Phys. Rev. D {\bf68} (2003) 064014, hep-th/0304060.

\bibitem{setelias2}
M. R. Setare and E. C. Vagenas, Phys. Lett. B {\bf584} (2004) 127, hep-th/0309092.

%%%%%%%%%%%%%%%%%%%%%%%%%%%

\bibitem{BBM}
V. Balasubramanian, J. de Boer and D. Minic, Phys. Rev. D {\bf65} (2002) 123508, hep-th/0110108.


%%%%%%%%%%%%%%%%%%%%%%%%%%%

\bibitem{medved1} A. J. M.  Medved, Phys. Rev. D {\bf66} (2002) 124009, hep-th/0207247.

%%%%%%%%%%%%%%%%%%%%%%%%%%%

\bibitem{cvetic}
M. Cveti$\check{c}$, S. Nojiri and S. D. Odintsov, Nucl. Phys. B {\bf628} (2002) 295, hep-th/0112045;
S. Nojiri and S. D. Odintsov, gr-qc/0112066.

%%%%%%%%%%%%%%%%%%%%%%%%%%

\bibitem{mignemi} S.~Mignemi, to appear in Phys. Rev. D, hep-th/0307205.

%%%%%%%%%%%%%%%%%%%%%%%%%%%%%%%%%


\bibitem{savonije}
I. Savonije and E. Verlinde, Phys. Lett. B {\bf507} (2001) 305, hep-th/0102042.


%%%%%%%%%%%%%%%%%%%%%%%%%%%%%%%%%%%%%%%

\bibitem{bounce}
S. Mukherji and M. Peloso, Phys. Lett. B {\bf547} (2002) 297, hep-th/0205180;
Y. Shtanov and V. Sahni, Phys. Lett. B {\bf557} (2003) 1, gr-qc/0208047.

%%%%%%%%%%%%%%%%%%%%%%%%%%%%%%%%%

\bibitem{shear}
R. Maartens, V. Sahni and T. D. Saini, Phys. Rev. D {\bf63} (2001) 063509, gr-qc/0011105;
A. V. Toporensky, Class. Quant. Grav. {\bf18} (2001) 2311, gr-qc/0103093.

%%%%%%%%%%%%%%%%%%%%%%%%%%%%%%%%%

\bibitem{stiff1}
A. K. Biswas and S. Mukherji, JHEP {\bf0103} (2001) 046, hep-th/0102138.

\bibitem{stiff2}
A. J. M.  Medved, hep-th/0111182.

%%%%%%%%%%%%%%%%%%%%%%%%%%%%%%%%%

\bibitem{gibbons}
G. W. Gibbons and S. W. Hawking, Phys. Rev. D {\bf15} (1977) 2752.

%%%%%%%%%%%%%%%%%%%%%%%%%%%%%%%%%

\bibitem{logcor}S. Nojiri, S. D. Odintsov and S. Ogushi, Int. J. Mod. Phys. A {\bf18} (2003) 3395, hep-th/0212047.

%%%%%%%%%%%%%%%%%%%%%%%%%%%%%%%%%

%%%%%%%%%%%%%%%%%%%%%%%%%%%%%%%%%%%%%%%%%%%%%%%%%%%%%%%%%%%%%
%%%%%%%%%%%%%%%%%%%%%%%%%%%%%%%%%%%%%%%%%%%%%%%%%%%%%%%%%%%%%
\end{thebibliography}
\end{document}